\documentclass[11pt]{article}
\usepackage{mathrsfs}
\usepackage{amsfonts}
\usepackage{layout}
\usepackage[top=2cm, bottom=2cm, left=2cm, right=2cm]{geometry}
\usepackage{amssymb}
\usepackage{amsmath}
\usepackage{amsthm}
\usepackage[colorlinks=true, linkcolor=blue, citecolor=red]{hyperref}
\makeatletter

\@addtoreset{equation}{section}
\makeatother

\newtheorem{corollary}{Corollary}[section]

\newtheorem{proposition}{Proposition}[section]
\newtheorem{remark}{Remark}[section]

\newtheorem{theorem}{Theorem}[section]

\begin{document}

\title{Closed-form expressions, generating function and Mehler-Heine type formula for the associated Meixner polynomials and some applications }
\author{Khalid Ahbli}
\maketitle
\begin{center}
Faculty of Sciences of Agadir, Ibn Zohr University, Morocco\\
E-mail: ahbli.khalid@gmail.com
\vspace*{0.2mm}\vspace*{0.2mm}
\end{center}

\begin{abstract} We give a closed-form expression for the associated Meixner polynomials from which we derive closed-form expressions for the associated Charlier and Laguerre polynomials by a limit procedure. These formulas are then used to derive generating functions of these polynomials. Also, we investigate the Mehler-Heine type asymptotic formulas for the associated Meixner and Charlier polynomials. Some applications and consequences of our results are also mentioned.
\end{abstract}

\section{Introduction}
Recall that, for a set of orthogonal polynomials $\left\{P_{n}\left( x\right) \right\} $ satisfying the three-term recurrence
relation 
\begin{equation}\label{3TRR}
\begin{split}
& A_{n}P_{n+1}(x)=\left( B_{n}\,x+C_{n}\right)
P_{n}(x)-D_{n}P_{n-1}(x),\quad n=1,2,3,... \\
& P_{-1}(x)=0,\text{ }P_{0}(x)=1
\end{split}
\end{equation}
where $A_{n},\,B_{n},\,C_{n},\,D_{n}$ are real coefficients such that 
\begin{equation}
A_{n-1}B_{n-1}B_{n}D_{n}>0,\quad n\geq 1,  \label{coef_cond}
\end{equation}%
the associated orthogonal polynomials (AOPs), denoted $\{P_{n}(x;\gamma )\}$, are defined by $%
\left( 1.1\right) $ in terms of coefficients $A_{n+\gamma }$, $B_{n+\gamma }$%
, $C_{n+\gamma }$ and $D_{n+\gamma }$ where $\gamma \geq 0$ is a parameter. They have recently generated wide interest and some work has been devoted to the task of obtaining explicit-form expressions (or closed-form expression) for these polynomials. This has been done successfully for the associated Jacobi polynomials \cite{Wimp1987} and their special cases, including the associated Gegenbauer \cite{Lew1993}, Laguerre and Hermite polynomials \cite{AW1984} and more recently for the associated Meixner-Pollaczek polynomials \cite{Luo2019}.  These polynomials have many applications in diverse fields, such as queuing and inventory models, chemical kinetics, population dynamics and quantum optics. For instance, it was shown in \cite{ILV_1988} that associated Meixner polynomials are birth and death process polynomials (a stationary Markov process) with rates $\lambda_n=c(n+\gamma+\beta)$, $\mu_n=n+\gamma$, $0<c<1$. In \cite{Ahbli2018,Ahbli2020}, associated Meixner-Pollaczek and Hermite polynomials were used to construct some new sets of nonlinear coherent states which play an important role in quantum optics. For other works on AOPs and their applications see the articles \cite{Hend1990, IR1991, BD1967, BI1982, ILVW1990(a), ILV_1988, ILVW1990(b), Rahman1996, Rahman2000, GIM1991, Lew1995, Darke2009, Wimp1990,Ahbli2018,Ahbli2020}, the book \cite{IsmailBook} and their references. 

One of the main topics in the theory of orthogonal polynomials is the
study of their asymptotics. Various types of asymptotics corresponding to orthogonal
polynomials can be studied, which provide valuable information on polynomials
with sufficiently large degrees. In particular, the Mehler-Heine formulas provide the
local asymptotics of the polynomials (a good approximations in the neighborhood of $x = 0$) and establish a limit
relationship between the polynomials and other types of (special) functions. Furthermore, they determine
the asymptotic limit of zeros of the polynomials as the degree $n$ goes to infinity \cite{Szego1939}. The classic Mehler-Heine formula was introduced by Heine \cite{Heine1861} and Mehler \cite{Mehler1868} in order to describe for the Legendre polynomials
of degree $n$ \ the following asymptotic behavior $\lim_{n\rightarrow
+\infty }P_{n}\left( \cos \left( \frac{z}{n}\right) \right) =J_{0}\left(
z\right) $ where $J_{\alpha }
$ is the Bessel function of the first kind and of order $\alpha$. This formula holds uniformly in every bounded domain in the complex plane. Its generalization to Jacobi polynomials (\cite[Sect.8.1]{Szego1939}) reads $\lim_{n\rightarrow +\infty }n^{-\alpha}P_{n}^{\left(
\alpha ,\beta \right) }\left( \cos \left( \frac{z}{n}\right) \right) =\left( 
\frac{z}{2}\right) ^{-\alpha }J_{\alpha }\left( z\right) $ and provides a complete
characterization of the function $P_{n}^{\left( \alpha ,\beta \right)
}\left( \cos \theta \right) $ for $\theta =O\left( n^{-1}\right) $. Consequently, it enables one to derive some properties of Bessel functions
from the corresponding properties of Jacobi polynomials and information about the distribution of the zeros of these polynomials (see \cite[Thm. 8.1.2, p.192]{Szego1939}). There are generalizations to other classical orthogonal polynomials, which are also called the Mehler-Heine formula or Meler-Heine type formula (see \cite{Dominici2016} for a review).

In this paper, we will be concerned with the associated Meixner polynomials (AMPs), denoted $\mathscr{M}_{n}(x;\beta
,c,\gamma )$, that correspond to coefficients $A_{n+\gamma }=c,\,B_{n+\gamma }=c-1,\,C_{n+\gamma }=(c+1)(n+\gamma )+\beta
c,\,D_{n+\gamma }=(n+\gamma )(n+\gamma +\beta -1)$ and reduce to the classical Meixner polynomials $M_n\left(x; \beta , c\right)$  \cite{Chihara} when $\gamma =0$. We prove that these polynomials are connected to another family of AOPs called the associated Meixner-Pollaczek polynomials (AMPPs) and denoted here by $\mathscr{P}_n^{(\nu)}(x;\varphi, \gamma )$. The later were obtained in \cite{Pollaczek1950} through the limit:
\begin{equation}\label{AP_to_AMP}
\mathscr{P}_n^{(\nu)}(x;\varphi, \gamma )=\lim\limits_{\epsilon \to 0}Q_{n}^{\nu}\left(\epsilon x+\cos \varphi ;\frac{\sin \varphi}{\epsilon} , -\frac{\cos \varphi \sin \varphi}{\epsilon}, \gamma\right)=Q_{n}^{\nu}(\cos \varphi;0,x \sin \varphi , \gamma)
\end{equation}
where $Q_{n}^{\nu}(y;a,b,\gamma)$ are the associated Pollaczek polynomials. When $\gamma=0$, we recover the classical Meixner-Pollaczek polynomials $P_n^{(\nu)}(x,\varphi)$. Recently, Luo and Raina \cite{Luo2019} have derived a closed-form expression as well as a generating function for $\mathscr{P}_n^{(\nu)}(x;\varphi, \gamma )$ ($P_n^{(\lambda)}(x;\varphi, c)$ in their notation) from the one of $Q_{n}^{\nu}(\cos \varphi ;a,b , \gamma)$ in view of the relation \eqref{AP_to_AMP}. We exploit their result and the relation connecting AMPPs and AMPs (see Eq.\eqref{AMPP_AMP} below) to derive a closed-form expression for the AMPs which after is used to write a generating function of these polynomials. Finally, we find a new representation of the AMPs from which the Mehler-Heine type formula for these polynomials can be obtained by directly letting $n\to\infty$. We believe that there are a few work in literature that treated the properties of the AMPs. This work is then coming to fill this gap.

The structure of the paper is as follows. In Section \ref{Sect_AMP}, we establish a closed-form expression and a generating function for the AMPs and mention several useful and interesting consequences of obtained results. In section \ref{Sect_ACP-ALP}, by using a limit procedure, we derive some closed-form expressions for the associated Laguerre and Charlier polynomials from the one of the AMPs which will be used to compute some generating functions of each of these two family of polynomials. Section \ref{Sect_MHF} is devoted to Mehler-Heine asymptotic formulas for associated Meixner and Charlier polynomials. In section \ref{proofs}, we give technical proofs. 

\section{The associated Meixner polynomials}\label{Sect_AMP}
%\subsection{A closed formula in terms of ${}_4F_3(1)$}
In this section, we first give a closed-form expression for the associated Meixner polynomials which is deduced from the one of the associated Meixner-Pollaczek polynomials obtained in \cite{Luo2019}. The key lemma here is the following connection relation (see section \ref{proofs} for proof):
\begin{equation}
\mathscr{P}_n^{(\nu)}(x;\varphi, \gamma )=\frac{e^{-in\varphi}}{(\gamma +1)_n}\mathscr{M}_n(ix-\nu ;2\nu,e^{-2i\varphi } ,\gamma ).\label{AMPP_AMP}
\end{equation}
 The obtained closed-form expression is then used to derive a generating function of these polynomials.
\subsection{Closed-form expression and Generating function}
The associated Meixner polynomials satisfy the recurrence relation:
\begin{equation}\label{AM_RR}
\begin{split}
c\mathscr{M}_{n+1}(x;\beta ,c,\gamma )=[(c-1)x+(c+1)(n+\gamma )& +\beta c]%
\mathscr{M}_{n}(x;\beta ,c,\gamma ) \\
& -(n+\gamma )(n+\gamma +\beta -1)\mathscr{M}_{n-1}(x;\beta ,c,\gamma ).
\end{split}
\end{equation}
When $\gamma =0$, we recover the recurrence relation of the classical Meixner polynomials defined by \cite{Chihara}:
\begin{equation}\label{Meixner_poly}
M_{n}(x;\beta, c)=(\beta)_n\, {}_2F{}_1\left( 
\begin{array}{c}
-n,\, -x \\ 
\beta
\end{array}
\left|1-\frac{1}{c}\right.\right).
\end{equation}
 Here, $(a)_{k}$ is the Pochhammer symbol defined by $(a)_{0}=1$, $(a)_{k}=a(a+1)\cdots (a+k-1)$, $k\geq 1$.
 
The next theorem gives a closed-form expression for the associated Meixner polynomials in terms of ${}_4F_3(1)$. We have been unable to find this result in the literature and therefore believe it to be new.
\begin{theorem}\label{thm_CF_AMP}
A closed-form expression for the associated Meixner
polynomials is given by
\begin{eqnarray}\label{AMP_4F3}
\begin{split}
\mathscr{M}_n(x;\beta , c, \gamma )=c^{-n}\frac{(\gamma +1 )_n(\gamma+\beta)_n}{n!}\sum_{k=0}^{n}(1-c)^k\frac{(-n)_k(\gamma+\beta+x)_k}{(\gamma +1 )_k(\gamma+\beta)_k}\\ 
\times {}_4F_3\left( 
\begin{array}{c}
k-n,\,\gamma+\beta+x+k ,\, \gamma+\beta -1,\, \gamma\ \\ 
 \gamma+\beta +x,\, \gamma+\beta +k, \, \gamma+1+k
\end{array}
\big| 1\right)
\end{split}
\end{eqnarray} 
with the conditions $c>0,\,c\neq 1$ and $\gamma +\beta >0$. When $\gamma =0$%
, we recover the Meixner polynomials \eqref{Meixner_poly}.
\end{theorem}
The closed-form expression \eqref{AMP_4F3} can also be written, using the relation \cite[Eq.(7)]{LRV_1996}: 
\begin{equation*}  \label{Rel_M-M}
\mathscr{M}_{n}(x;\beta, c,\gamma)=c^{-n}\mathscr{M}_{n}(-\beta-x;\beta,
c^{-1},\gamma),
\end{equation*}
which follows directly from the recurrence relation, as 
\begin{eqnarray}\label{AMP_4F3_2}
\begin{split}
\mathscr{M}_n(x;\beta , c, \gamma )=\frac{(\gamma +1 )_n(\gamma+\beta)_n}{n!}\sum_{k=0}^{n}\tilde{c}^k\frac{(-n)_k(\gamma-x)_k}{(\gamma +1 )_k(\gamma+\beta)_k} {}_4F_3\left( 
\begin{array}{c}
k-n,\,\gamma-x+k ,\, \gamma+\beta -1,\, \gamma\ \\ 
 \gamma-x,\, \gamma+\beta +k, \, \gamma+1+k
\end{array}
\big| 1\right)
\end{split}
\end{eqnarray}
where $\tilde{c}=\frac{c-1}{c}$. 
%\subsection{A generating function}

Now, we give a generating function for the associated Meixner polynomials.
\begin{theorem}\label{thm_GF_AMP}
A generating function for the associated Meixner polynomials is given by 
\begin{eqnarray}\label{gen2_AMP}
\sum_{n=0}^{+\infty}\frac{(c\,t)^n}{(\gamma+\beta)_n}\mathscr{M}_n(x;\beta , c, \gamma )=\left(1-c\,t\right)^{-1}F_1\left[1 , \gamma , -x ; \gamma +\beta ; t, \frac{t(1-c)}{1-ct} \right]
\end{eqnarray} 
for $|t|<\min \{1,\, c^{-1}\}$, in terms of Appel hypergeometric function $F_1$ defined by \eqref{Appel_F1}.\\ In particular, we have
\begin{eqnarray}\label{gener2_MPP}
\sum_{n=0}^{+\infty}M_n(x;\beta , c)t^n=\left(1-t\right)^{-1}{}_2F_1\left( 
\begin{array}{c}
1,\, -x \ \\ 
\beta
\end{array}
\left| \frac{t(1-c)}{c(1-t)}\right.\right).
\end{eqnarray} 
\end{theorem} 

\noindent Another generating function for these polynomials can be deduced from \eqref{AMPP_AMP} and the generating function for the associated Meixner-Pollaczek polynomials given in \cite[p.3(1.7)]{Luo2019} as 
\begin{eqnarray}\label{gen_AMP}
\sum_{n=0}^{+\infty}\frac{(c\,t)^n}{(\gamma+1)_n}\mathscr{M}_n(x;\beta , c, \gamma )=\left(1-c\,t\right)^{-\beta-x}\left(1-t\right)^{x}F_1\left[\gamma , 1-\beta-x, 1+x ; \gamma +1 ; c\,t, t \right].
\end{eqnarray} 
\noindent Letting $c=1$ in \eqref{gen_AMP} and applying the following reduction formula for $F_1$ \cite[p.79(1)]{Bailey1964}:
\begin{eqnarray}
F_1\left[\alpha_1 ; \,\lambda_1 ,\, \lambda_2 ;\,\alpha_2 ;\, t,\, t \right]={}_2F_1\left( 
\begin{array}{c}
\alpha_1 ,\, \lambda_1 +\lambda_2\ \\ 
 \alpha_2
\end{array}
\big| t\right),
\end{eqnarray}
for $\alpha_1=\gamma,\, \lambda_1=1-\beta-x,\, \lambda_2=1+x$ and $\alpha_2=\gamma+1$, gives 
\begin{eqnarray}
\sum_{n=0}^{+\infty}\frac{(\gamma+\beta)_n }{n!} {}_3F_2\left( 
\begin{array}{c}
-n ,\, \gamma+\beta-1,\, \gamma\ \\ 
 \gamma+\beta ,\, \gamma+1
\end{array}
\big| 1\right)\,t^n=(1-t)^{-\beta}{}_2F_1\left( 
\begin{array}{c}
2-\beta,\, \gamma\ \\ 
 \gamma+1
\end{array}
\big| t\right),\quad |t|<1.
\end{eqnarray}
which can be identifies as a special case of the formula given in \cite[p.107(14)]{Sr-Ma1984}.
Again, if we take $\gamma=0$ in \eqref{gen_AMP}, we get
\begin{eqnarray}\label{gener_MPP}
\sum_{n=0}^{+\infty}\frac{(c\,t)^n}{n!}M_n(x;\beta , c)=\left(1-c\,t\right)^{-\beta-x}\left(1-t\right)^{x}.
\end{eqnarray} 
The integral representation of the generating function \eqref{gen_AMP} was obtained in \cite[p.345(4.4)]{ILV_1988} as
\begin{equation}\label{int_repr_gener_AMP}
\sum_{n=0}^{+\infty}\frac{(c\,t)^n}{(\gamma+1)_n}\mathscr{M}_n(x;\beta , c, \gamma )=\gamma\left(1-c\,t\right)^{-\beta-x}\left(1-t\right)^{x}\int_{0}^{1}u^{\gamma -1}\left(1-c\,t\,u\right)^{x+\beta-1}\left(1-t\,u\right)^{-x-1}d\,u ,
\end{equation}
where the integral in the RHS of \eqref{int_repr_gener_AMP} is nothing than the integral representation of the Appel function $F_1$ \cite[p.77(4)]{Bailey1964}. An expression of $\mathscr{M}_n(x;\beta , c, \gamma )$ in terms of $M_{n}(x;\beta , c)$ follows as a simple consequence of \eqref{gener_MPP}-\eqref{int_repr_gener_AMP}. Precisely, 
\begin{corollary} We have the relation
\begin{equation}
\frac{n!\, \mathscr{M}_n(x;\beta , c, \gamma )}{(\gamma+1)_n}=\sum_{k=0}^{n} \displaystyle \binom{n}{k} \frac{\gamma }{(k+\gamma)}M_{n-k}(x;\beta , c)M_{k}(-x-1;2-\beta , c).
\end{equation} 
\noindent Moreover, we have the following generating function
\begin{equation}\label{gen_MP}
\sum_{n=0}^{+\infty} \frac{\gamma \,(c\,t)^n}{(n+\gamma)n!}M_{n}(x;\beta , c)=F_1\left[\gamma , x+\beta , -x ; \gamma +1 ; c\,t, t \right],\quad |t|< |c|^{-1},\, (\gamma \text{ arbitrary}).
\end{equation}
\end{corollary}
\subsection{Some identities}
The only expression we find in literature for these polynomials was given in a quadratic form made out of Gauss hypergeometric functions  ${}_2F_1$ (see \cite[Eq(21)]{LRV_1996}) as 
\begin{eqnarray}\label{m_n}
\begin{split}
\mathscr{M}_n(x;\beta ,c,\gamma)=\frac{1}{\beta -1}\left\{ (\gamma +\beta -1)_{n+1} \, {}_2F_1\left( 
\begin{array}{c}
x+1,\gamma \\ 
2-\beta
\end{array}
\big|\tilde{c}\right){}_2F_1\left( 
\begin{array}{c}
-x,-n-\gamma \\ 
\beta
\end{array}
\big|\tilde{c}\right)\right.\\
\left. -(\gamma)_{n+1}\,{}_2F_1\left( 
\begin{array}{c}
x+\beta,\gamma+\beta -1 \\ 
\beta
\end{array}
\big|\tilde{c}\right){}_2F_1\left( 
\begin{array}{c}
1-\beta-x,-n-\gamma-\beta +1 \\ 
2-\beta
\end{array}
\big|\tilde{c}\right)\right\}
\end{split}
\end{eqnarray}
where $\tilde{c}=\frac{c-1}{c}$. This
relation is valid under restrictions $\gamma >0$, $\gamma +\beta >0$, $%
\gamma +\beta \neq 1$, and $\beta \neq 1,2,...$ . By combining \eqref{AMP_4F3} and \eqref{m_n} then making slight modifications of notations, we can readily obtain the following finite sum formula for ${}_4F_3(1)$: 
\begin{corollary}
The following finite sum
\begin{eqnarray}  \label{sum_fin_4F3}
\begin{split}
&\sum_{k=0}^{n}t^k \frac{(-n)_k(a+y)_k}{(a +1 )_k(b+1)_k}{}_4F_3\left( 
\begin{array}{c}
k-n,\,a+y+k ,\, a,\, b \  \\ 
a+y,\, b+1 +k, \, a+1+k%
\end{array}
\big| 1\right)& \\
&=\frac{n!}{b-a}\left\{ \frac{b}{(a +1 )_n} \, {}_2F_1\left( 
\begin{array}{c}
1-y,a \\ 
a-b+1%
\end{array}
\big|t\right){}_2F_1\left( 
\begin{array}{c}
y,-n-a \\ 
b-a+1%
\end{array}
\big|t\right)\right. & \\
&\left. -(1-t)^{n+1}\frac{a}{(b+1)_n}\,{}_2F_1\left( 
\begin{array}{c}
1-y,n+a +1 \\ 
a-b+1%
\end{array}
\big|t\right){}_2F_1\left( 
\begin{array}{c}
y,1-a \\ 
b-a+1%
\end{array}
\big|t\right)\right\} &
\end{split}%
\end{eqnarray}
is valid for $a>0,\, b>-1,\, b\neq 0$ and $b-a\neq 0,1,2,...$ .
\end{corollary}

\noindent Two obvious particular cases of \eqref{sum_fin_4F3} are 
\begin{equation}
{}_{3}F_{2}\left( 
\begin{array}{c}
-n,\,a,\,b\  \\ 
a+1,\,b+1%
\end{array}%
\big|1\right) =\frac{n!}{b-a}\left\{ \frac{b}{(a+1)_{n}}-\frac{a}{(b+1)_{n}}%
\right\} ,  \label{3F2_pocha}
\end{equation}%
and 
\begin{equation*}
\begin{split}
\sum_{k=0}^{n}t^{k}\frac{(-n)_{k}}{(b+1)_{k}}{}_{3}F_{2}\left( 
\begin{array}{c}
k-n,\,a,\,b\  \\ 
a+1,\,b+1+k%
\end{array}%
\big|1\right) =& \frac{n!}{b-a}\left\{ \frac{b}{(a+1)_{n}}%
\,{}_{2}F_{1}\left( 
\begin{array}{c}
1,-n-a \\ 
b-a+1%
\end{array}%
\big|t\right) \right. \\
& \left. -(1-t)^{n+1}\frac{a}{(b+1)_{n}}\,{}_{2}F_{1}\left( 
\begin{array}{c}
1,1-a \\ 
b-a+1%
\end{array}%
\big|t\right) \right\} .
\end{split}%
\end{equation*}%
It should be noted here that a generalization of \eqref{3F2_pocha} for a
positive integer $m$ in denominator parameter was proved in \cite[p.37(5.7)]{MP2012} and is given by 
\begin{equation*}
{}_{3}F_{2}\left( 
\begin{array}{c}
-n,\,a,\,b\  \\ 
a+m,\,b+1%
\end{array}%
\big|1\right) =\frac{(a)_{m}}{(a-b)_{m}}\left\{ \frac{(1)_{n}}{(1+b)_{n}}-%
\frac{b}{a}\sum\limits_{l=0}^{m-1}\frac{(a-b)_{l}(1+l)_{n}}{%
(1+a)_{l}(1+a+l)_{n}}\right\} .
\end{equation*}%
We note also that if we put $c=1$ in \eqref{AMP_4F3} then, the obtained
expression 
\begin{equation*}
\mathscr{M}_{n}(x;\beta ,1,\gamma )=\frac{(\gamma +1)_{n}(\gamma +\beta )_{n}%
}{n!}{}_{3}F_{2}\left( 
\begin{array}{c}
-n,\,\gamma +\beta -1,\,\gamma \  \\ 
\gamma +\beta ,\,\gamma +1%
\end{array}%
\big|1\right)=\frac{1}{\beta-1}\{(\gamma+\beta-1)_{n+1}-(\gamma )_{n+1}\}  
\end{equation*}%
do not depend on $x$. That is $\mathscr{M}_{n}(x;\beta ,1,\gamma )$ is not a
polynomials in $x$ but it still satisfies \eqref{AM_RR}. This means that if
we have two representations of a given family of orthogonal polynomials one
may deduce some useful formulas when we broke the condition \eqref{coef_cond}%
.
\section{The associated Charlier and Laguerre Polynomials}\label{Sect_ACP-ALP}
In this section, we exploit the result obtained in the previous section to derive closed-form expressions for the associated Charlier and Laguerre polynomials by using a limit procedure. Also, we give some generating functions of each of these polynomials.
\subsection{The associated Charlier Polynomials}
By comparing the recurrence relations of the associated Meixner and Charlier
polynomials, it is easy to check the limiting relation (see \cite[%
Eq.(35)]{LRV_1996}): 
\begin{equation}
\mathscr{C}_{n}(x;a,\gamma )=\lim\limits_{\beta \rightarrow \infty }\frac{1}{%
(\gamma +\beta )_{n}}\mathscr{M}_{n}\left( x;\beta ,\frac{a}{a+\beta }%
,\gamma \right).  \label{LR_CM}
\end{equation}%
Thanks to this relation we can obtain from \eqref{AMP_4F3_2} and by using the limit 
\begin{equation}  \label{limit_pocha}
\lim\limits_{a\to \infty}\frac{(ax)^k}{(ay+b)_k}=\left(\frac{x}{y}\right)^k,
\ k\geq 1,
\end{equation}
where $x,y,b$ are fixed, a closed-form expression for
associated Charlier polynomials as follows
\begin{eqnarray}  \label{ACP_exp}
\mathscr{C}_n(x;a,\gamma )=\frac{(\gamma +1 )_n}{n!}\sum_{k=0}^{n}(-a)^{-k}%
\frac{(-n)_k(\gamma-x)_k}{(\gamma +1 )_k} {}_3F_2\left( 
\begin{array}{c}
k-n,\,\gamma-x+k ,\, \gamma\  \\ 
\gamma-x,\, \gamma+k+1%
\end{array}
\big| 1\right)
\end{eqnarray}
with the condition $a> 0$. When the following ${}_3F{}_2(1)$ transformation formula \cite[p.142]{AAR1999}:
\begin{eqnarray}  \label{3F2_trans}
{}_3F{}_2\left( 
\begin{array}{c}
-m,\,a,\,b \\ 
c,\,d%
\end{array}
\big|1\right)=\frac{(c-a)_m}{(c)_m} {}_3F_2\left( 
\begin{array}{c}
-m,\,a,\,d-b \\ 
a-c+1-m,\, d%
\end{array}
\big|1\right),
\end{eqnarray}
is used, the closed-form expression \eqref{ACP_exp} becomes 
\begin{eqnarray}  \label{ACP_exp2}
\mathscr{C}_n(x;a,\gamma )=\sum_{k=0}^{n}(-a)^{-k}\frac{(-n)_k(\gamma-x)_k }{k!}
\ {}_3F_2\left( 
\begin{array}{c}
-k,\,\gamma ,\, k-n \  \\ 
-n,\, \gamma-x%
\end{array}
\big| 1\right).
\end{eqnarray}
%\subsection{A generating function for associated Charlier polynomials}
Following the same lines as for the associated Meixner polynomials, we give the generating function for the associated Charlier polynomials and some consequences. Precisely, we have the following result.
\begin{proposition}\label{prop_GF_ACP}
A generating function for the associated Charlier polynomials is given by
\begin{eqnarray}\label{gener_ACP}
\sum_{n=0}^{+\infty}\frac{t^n}{(\gamma+1)_n}\mathscr{C}_n(x;a, \gamma )=e^{t}\left(1-\frac{t}{a}\right)^{x}\Phi_1\left[\gamma , x+1 ; \gamma +1 ; \frac{t}{a}, -t \right], \quad |t|<|a|,
\end{eqnarray} 
where $\Phi_1$ is the Humbert's confluent hypergeometric function defined by \eqref{Phi_1F1}.\\
In particular, we have
\begin{eqnarray}\label{gener_CP}
\sum_{n=0}^{+\infty}\frac{t^n}{n!}C_n(x;a)=e^{t}\left(1-\frac{t}{a}\right)^{x}.
\end{eqnarray} 
\end{proposition}

The proof of \eqref{gener_ACP} is direct (see Sect. \ref{proofs}). An alternative method to obtain generating function is to start from the recurrence relation of the polynomials and find the differential equation satisfied by this generating function which now must be resolved. In our case, if we denote the LHS of \eqref{gener_ACP} by $\mathscr{G}(x,t)$, the recurrence relation of $\mathscr{C}_n(x;a, \gamma )$ leads to
\begin{equation}\label{equ-diff-ACP}
t(a-t)\frac{\partial}{\partial t}\mathscr{G}(x,t)+\left[t^2+(x-a-\gamma )t+a\gamma \right]\mathscr{G}(x,t)=a\gamma .
\end{equation}
We make the substitution
\begin{equation}\label{subst}
\mathscr{G}(x,t)=e^{t}\left(1-\frac{t}{a}\right)^{x}\mathscr{H}(x,t)
\end{equation}
and obtain for $\mathscr{H}(x,t)$ the following equation:
\begin{equation}
t\frac{\partial}{\partial t}\mathscr{H}(x,t)+\gamma \mathscr{H}(x,t)=\gamma e^{-t}\left(1-\frac{t}{a}\right)^{-x-1}.
\end{equation}
The solution of the above equation, after taking into account the condition $\mathscr{G}(x,0)=1$, is 
\begin{equation}
\mathscr{H}(x,t)=\gamma \int_{0}^{1}u^{\gamma-1}e^{-u\,t}\left(1-\frac{u\,t}{a}\right)^{-x-1}du, \quad |t|<\min \{1,\, |a|\}.
\end{equation}
Thus, $\mathscr{G}(x,t)$ has the following integral representation 
\begin{equation}\label{int_repr_GF_ACP}
\mathscr{G}(x,t)=\gamma  e^{t}\left(1-\frac{t}{a}\right)^{x}\int_{0}^{1}u^{\gamma-1}e^{-t\,u}\left(1-\frac{t\,u}{a}\right)^{-x-1}du.
\end{equation}
Note that in order to simplify the differential equation \eqref{equ-diff-ACP}, we have made the substitution \eqref{subst} which is nothing than writing the generating function of the associated polynomials as a product of the (analogous) generating function of (corresponding) classical polynomials and another function which is now easy to find. We believe that this is the best substitution we may perform when we look for the generating function of associated orthogonal polynomials.  

An easy consequence of the representation \eqref{int_repr_GF_ACP} and \eqref{gener_CP} is 
\begin{corollary}
\begin{equation}
\frac{n!\,\mathscr{C}_n(x;a, \gamma )}{(\gamma+1)_n}=\sum\limits_{k=0}^{n}\displaystyle \binom{n}{k}\frac{\gamma(-1)^k}{(\gamma +k)}C_{n-k}(x;a)C_k(-x-1;-a).
\end{equation}
Consequently, we get the following generating function
\begin{eqnarray}\label{gener_2_CP}
\sum_{n=0}^{+\infty}\frac{\gamma}{\gamma+n}C_n(x;a)\frac{t^n}{n!}=\Phi_1\left[\gamma , -x ; \gamma +1 ; \frac{t}{a}, t \right], \quad |t|<|a|, \, (\gamma \text{ arbitrary}).
\end{eqnarray} 
\end{corollary}
\subsection{The associated Laguerre Polynomials}
By comparing the recurrence relations of the associated Meixner polynomials and the associated Laguerre polynomials, it is easy to check the relation (see \cite[Eq.(39)]{LRV_1996}): 
\begin{equation}
\mathscr{L}_{n}^{(\alpha )}(x;\gamma )=\lim\limits_{c\rightarrow 1}\frac{1}{%
(\gamma +1)_{n}}\mathscr{M}_{n}\left( \frac{x}{1-c};\alpha +1,c,\gamma
\right) .  \label{LR_ALP-AMP}
\end{equation}%
This limit relation allow us to write an explicit formula for the associated Laguerre polynomials as follows
\begin{equation}  \label{ALP_exp}
\mathscr{L}_n^{(\alpha)}(x;\gamma )=\frac{(\gamma +\alpha+1 )_n}{n!}%
\sum_{k=0}^{n}\frac{(-n)_k \, x^k}{(\gamma +1 )_k(\gamma +\alpha+1 )_k} {}%
_3F_2\left( 
\begin{array}{c}
k-n,\,\gamma+\alpha ,\, \gamma\  \\ 
\gamma+\alpha+k+1,\, \gamma+k+1%
\end{array}
\big| 1\right)
\end{equation}
where $\alpha >-1$.
\begin{remark}
By a different method, the explicit polynomial form \eqref{ALP_exp}
was found in \cite[p.22, Eq(2.8)]{AW1984}. It can also be obtained from the associated Meixner-Pollaczek
polynomials by using the limit relation: 
\begin{equation*}
\mathscr{L}_n^{(\alpha)}(x;\gamma )=\lim\limits_{\varphi\to 0}\mathscr{P}%
_n^{((\alpha+1)/2)}\left(\frac{-x}{2\sin\varphi};\,\varphi,\, \gamma \right),
\end{equation*}
as was explained by Rahman in \cite[p.7]{Rahman2000}. Using the transformation %
\eqref{3F2_trans}, these polynomials can be rewritten as
follows (see also \cite[Eq(1.34)]{Rahman2000}): 
\begin{equation}
\mathscr{L}_n^{(\alpha)}(x;\gamma )=\frac{(\alpha+1 )_n}{n!}\sum_{k=0}^{n}%
\frac{(-n)_k \, x^k}{(\gamma +1 )_k(\alpha+1 )_k} {}_3F_2\left( 
\begin{array}{c}
k-n,\,1-\alpha+k ,\, \gamma\  \\ 
-\alpha-n,\, \gamma+k+1%
\end{array}
\big| 1\right).
\end{equation}
\end{remark}
We now give a generating function for the associated Laguerre polynomials. It was already found in \cite[p.25]{AW1984} and was derived there by using the method of differential equations. The proof we present here (see Sect. \ref{proofs} below) is different, direct and simpler.
\begin{proposition}\label{prop_GF_ALP}
A generating function of the associated Laguerre polynomials is 
\begin{eqnarray}\label{gen_ALP_1}
\sum_{n=0}^{+\infty}t^n\mathscr{L}_n^{(\alpha)}(x;\gamma )=(1-t)^{-\gamma-\alpha-1}\exp\left(\frac{x\,t}{t-1}\right)\Phi_1 \left[\gamma,\,\gamma+\alpha,\, \gamma+1;\, \frac{t}{t-1};\, \frac{-x\,t}{t-1} \right]
\end{eqnarray}
where $|t|<\frac{1}{2} ,\, x\in\mathbb{R}$ and $\Phi_1$ is the Humbert confluent hypergeometric function defined by \eqref{Phi_1F1}.\\ In particular, for $\gamma=0$, we recover the generating function of classical Laguerre polynomials
\begin{eqnarray}\label{gener_LP}
\sum_{n=0}^{+\infty}t^n L_n^{(\alpha)}(x)=(1-t)^{-\alpha-1}\exp\left( \frac{x\,t}{t-1} \right),\quad |t|<1 ,\, |x|<+\infty .
\end{eqnarray}
\end{proposition}
Using the connection formula \cite[p.24(2.20)]{AW1984}:
\begin{equation}
\mathscr{L}_n^{(\alpha)}(x;\gamma )=\sum_{k=0}^{n}\frac{\gamma}{k+\gamma}L_{n-k}^{(\alpha)}(x)L_k^{(-\alpha)}(-x),
\end{equation}
one can prove the following generating function for Laguerre polynomials
\begin{eqnarray}\label{Cor_gener_ALP_1}
\sum_{n=0}^{+\infty} \frac{\gamma}{n+\gamma}L_n^{(\alpha)}(x)t^n=(1-t)^{-\gamma}\Phi_1 \left[\gamma,\,\gamma -\alpha,\, \gamma+1;\, \frac{t}{t-1};\, \frac{x\,t}{t-1} \right]\quad (\gamma \text{ arbitrary}).
\end{eqnarray}
In particular, when $\alpha =\gamma$, we have (see also \cite[Eq(9.12.12)]{askey-scheme})
\begin{equation}\label{par_case}
\sum_{n=0}^{+\infty} \frac{\alpha}{n+\alpha}L_n^{(\alpha)}(x)t^n=(1-t)^{-\alpha} {}_1F_1\left( 
\begin{array}{c}
\alpha \ \\ 
 \alpha +1
\end{array}
\left| \frac{x\,t}{t-1}\right. \right).
\end{equation}
It is easy to see that \eqref{gener_LP} follows if we multiply \eqref{par_case} by $t^\alpha$ then differentiate with respect to $t$ and use the recurrence relation of confluent hypergeometric function \cite[p.325(13.3.4)]{NIST}.
\begin{remark}
Notice that the obtained generating functions for the associated Laguerre and Charlier polynomials may be reached from the one of the associated Meixner polynomials through limiting procedure by using the relations \eqref{LR_CM}, \eqref{LR_ALP-AMP} and 
the limits
\begin{eqnarray*}
e^z=\lim\limits_{|\mu|\to \infty}\left(1-\frac{z}{\mu}\right)^{-\mu} \quad \text{and}\quad \Phi_1\left[\alpha_1 ; \,\lambda ;\,\alpha_2 ;\, x,\, y \right]=\lim\limits_{|\mu |\to \infty}F_1\left[\alpha_1 ; \,\lambda ,\, \mu ;\,\alpha_2 ;\, x,\, \frac{y}{\mu} \right].
\end{eqnarray*}
%and
%\begin{eqnarray}
%\Phi_1\left[\alpha_1 ; \,\lambda ;\,\alpha_2 ;\, x,\, y \right]&=&\lim\limits_{|\mu |\to \infty}F_1\left[\alpha_1 ; \,\lambda ,\, \mu ;\,\alpha_2 ;\, x,\, \frac{y}{\mu} \right].
%\end{eqnarray}
Some transformations of Appel hypergeometric function $F_1$ are also needed (see \cite[p.78]{Bailey1964}).
\end{remark}

\section{Mehler-Heine type formula}\label{Sect_MHF}
To prove the Mehler-Heine type formula for classical orthogonal polynomials usually
the explicit-form expressions of these polynomials are used (see \cite[Sect.8.1]{Szego1939}). Similar explicit-form expression was obtained by Askey and Wimp for the associated Laguerre polynomials and used to prove Mehler-Heine type formula \cite[p.22]{AW1984} for these polynomials. But, the closed-form expression \eqref{AMP_4F3} is so complicated that it is difficult
to apply the same method as classical polynomials to our case. Therefore, we
shall give new different representation of $\mathscr{M}_n(x;\beta , c, \gamma )$ in terms of Gauss hypergeometric function $_2F_1$. Then, we shall prove the desired Mehler-Heine type formula from this representation by using asymtotics formulas for Gauss hypergeometric function. The obtained representation is derived from \eqref{m_n} and look like it. However, the drawback of \eqref{m_n} is that $n$ appears only in one of the first two parameters in $_2F_1$. Therefore, letting $n\to \infty$ would not lead us to an asymptotic formula directly.

Before stating the Mehler-Heine type formula for the associated Meixner polynomials, we show that $\mathscr{M}_n(x;\beta ,c,\gamma)$ can be expressed as a cross products of Gauss hypergeometric functions. More precisely, we have

\begin{proposition}\label{AMP_QF} For $x\in\mathbb{C}$, the associated Meixner polynomial can be expressed as
\begin{eqnarray}\label{HR_AMP}
\begin{split}
\mathscr{M}_n(x;\beta ,c,\gamma)=(1-c)^{1-\beta}\left[c^{-n}(\gamma -x)_{n} \mathsf{F}_{n+1}(c)\mathsf{G}_0(c)- \frac{c\ (\gamma)_{n+1}(\gamma+\beta-1)_{n+1}}{(\gamma-x-1)_{n+2}}  \mathsf{F}_{0}(c)\mathsf{G}_{n+1}(c)\right]
\end{split}
\end{eqnarray}
where
\begin{eqnarray}
\mathsf{F}_n(c):={}_2F{}_1\left( 
\begin{array}{c}
x+1 ,\, 2-\beta-\gamma-n \\ 
2+x-\gamma-n
\end{array}
\left|c\right.\right), \quad \mathsf{G}_n(c):={}_2F{}_1\left( 
\begin{array}{c}
\gamma+n ,\, 1-\beta-x \\ 
\gamma -x+n
\end{array}
\left|c\right.\right).
\end{eqnarray}
\end{proposition}
The above result is the departure point to establish the Mehler-Heine type formula for the associated Meixner polynomials. To do this, we start by writing
\begin{eqnarray}
\begin{split}
\frac{c^n}{(\gamma -x)_{n}}\mathscr{M}_n(x;\beta ,c,\gamma)= (1-c)^{1-\beta}\left[\mathsf{F}_{n+1}(c)\mathsf{G}_0(c)- c^{n+1}\frac{(\gamma)_{n+1}(\gamma+\beta-1)_{n+1}}{(\gamma -x)_{n}(\gamma-x-1)_{n+2}}  \mathsf{F}_{0}(c)\mathsf{G}_{n+1}(c)\right].
\end{split}
\end{eqnarray}
Next, we use the following asymptotic expansions of the hypergeometric function ${}_2F{}_1$ \cite{Cvi2017}:
\begin{equation}
{}_2F{}_1\left( 
\begin{array}{c}
a+\varepsilon \lambda ,\, b \\ 
c+\lambda
\end{array}
\big|z\right)\thicksim (1-\varepsilon z)^{-b}\quad \text{ as } |\lambda |\to \infty ,
\end{equation}
valid for $|\varepsilon |\leq 1$, and a ratio of Gamma functions \cite[p.24]{Sr-Ma1984}
\begin{equation}
\frac{\Gamma(z+a)}{\Gamma(z+b)}\thicksim z^{a-b} \text{ as } |z|\to \infty ,
\end{equation}
where $a$ and $b$ are bounded complex numbers, to show that
\begin{equation}
\mathsf{F}_{n+1}(c) \thicksim (1-c)^{-1 -x}  , \quad \mathsf{G}_{n+1}(c) \thicksim (1-c)^{x+\beta-1} 
\end{equation}
 and 
\begin{eqnarray}
\frac{(\gamma)_{n+1}(\gamma+\beta-1)_{n+1}}{(\gamma -x)_{n}(\gamma-x-1)_{n+2}} \thicksim n^{2x+\beta} \frac{\Gamma(\gamma -x)\Gamma(\gamma-x-1)}{\Gamma(\gamma)\Gamma(\gamma+\beta-1)}
\end{eqnarray}
as $n\to \infty $. Therefore, we have
\begin{eqnarray}\label{asym_meixner}
\begin{split}
\frac{c^n}{(\gamma -x)_{n}}\mathscr{M}_n(x;\beta ,c,\gamma)&\thicksim (1-c)^{-\beta -x} {}_2F{}_1\left( 
\begin{array}{c}
\gamma ,\, 1-\beta-x \\ 
\gamma -x
\end{array}
\left|c\right.\right)&\\
&- n^{2x+\beta}c^{n+1}\left(1-c\right)^x {}_2F{}_1\left( 
\begin{array}{c}
x+1 ,\, 2-\beta-\gamma \\ 
2+x-\gamma
\end{array}
\left|c\right.\right) \text{ as } n\to \infty .
&
\end{split}
\end{eqnarray}
For $0<b<1$ and bounded $a$, we can easily verify that $\lim\limits_{n\to +\infty}n^ab^n=0$. Thus, for $0<c<1$ we can omit the second term in \eqref{asym_meixner} to finally get the following result.
\begin{theorem} Let $0<c<1$. Then, for all bounded complex numbers $x$, we have
\begin{eqnarray}
\begin{split}
\lim\limits_{n\to \infty}\frac{c^n}{\Gamma(n+\gamma -x)}\mathscr{M}_n(x;\beta ,c,\gamma)=\frac{(1-c)^{-\beta -x}}{\Gamma(\gamma -x)} {}_2F{}_1\left( 
\begin{array}{c}
\gamma ,\, 1-\beta-x \\ 
\gamma -x
\end{array}
\left|c\right.\right).
\end{split}
\end{eqnarray}
In particular, for $\gamma=0$, we have from \cite[Equ.(29)]{Dominici2016}
\begin{equation}
\lim\limits_{n\to \infty}\frac{c^n}{\Gamma(n-x)}M_{n}(x;\beta, c)=\frac{(1-c)^{-\beta-x}}{\Gamma(-x)}.
\end{equation}
\end{theorem}
We precise here that the definition of Meixner polynomials used in this work differ from the one given in \cite{Dominici2016} by $(\beta)_n$. Now, we use the closed-form expression of the associated Charlier polynomials \eqref{ACP_exp2} to establish the Mehler-Heine type formula for these polynomials.
\begin{theorem}\label{thm_MH_ACP} For all bounded complex numbers $x$, we have
\begin{equation}\label{MH-ACP} 
\lim\limits_{n\to \infty}\frac{a^n}{\Gamma(n+\gamma-x)}\mathscr{C}_n(x;a,\gamma )=\frac{e^a}{\Gamma(\gamma-x)} {}_1F_1(\gamma;\ \gamma -x; -a).
\end{equation}
For $\gamma=0$, we have from \cite[Equ.(13)]{Dominici2016}
\begin{equation}
\lim\limits_{n\to \infty}\frac{a^n}{\Gamma(n-z)}C_n(z;a)=\frac{e^a}{\Gamma(-z)}.
\end{equation} 
\end{theorem}
Notice that Mehler-Heine formula of Charlier polynomials can be deduced from the one of Meixner polynomials by the limit relation obtained from \eqref{LR_CM} by taking $\gamma=0$. This remains also true for the associated Charlier polynomials.
\begin{remark}
The hypergeometric representation \eqref{HR_AMP} for the associated Meixner polynomials can also be obtained by using the connection relation \eqref{AMPP_AMP} and the hypergeometric representation for associated Meixner-Pollaczek polynomials 
\begin{equation}
\mathscr{P}_n^{(\nu)}(x;\varphi, \gamma )=e^{in\varphi}\frac{(\nu-ix+\gamma)_n}{(\gamma+1)_n}\tilde{\mathsf{F}}_{n+1}(\varphi)\tilde{\mathsf{G}}_{0}(\varphi)-\gamma e^{-i(n+2)\varphi}\frac{(2\nu+\gamma-1)_{n+1}}{(\nu-ix+\gamma-1)_{n+2}}\tilde{\mathsf{F}}_{0}(\varphi)\tilde{\mathsf{G}}_{n+1}(\varphi)
\end{equation}
where
\begin{eqnarray}
\tilde{\mathsf{F}}_{n}(\varphi):={}_2F{}_1\left( 
\begin{array}{c}
1-\gamma-n ,\, \nu+ix \\ 
2-\gamma-n-\nu+ix
\end{array}
\left|e^{-2i\varphi}\right.\right),\quad \tilde{\mathsf{G}}_{n}(\varphi):={}_2F{}_1\left( 
\begin{array}{c}
n+\gamma ,\, 1-\nu-ix \\ 
\gamma+n+\nu-ix
\end{array}
\left|e^{-2i\varphi}\right.\right),
\end{eqnarray}
which can be deduced from the hypergeometric representation for the associated pollaczek polynomials given in \cite[Theorem 2.1]{Luo2018} by means of the relation \eqref{AP_to_AMP}.
\end{remark}
\begin{remark}
In \cite{ILV_1988}, Ismail et al. proved the following asymptotic formula for the associated Meixner polynomials
\begin{equation*}
n^{x+1}\mathscr{M}_{n}(x;\beta ,c,\gamma )\thicksim (1-c)^{-\beta -x}\frac{%
\Gamma (\gamma +1)}{\Gamma (\gamma -x)}{}_{2}F{}_{1}\left( 
\begin{array}{c}
\gamma ,\,1-\beta -x \\ 
\gamma -x%
\end{array}%
\left\vert c\right. \right) ,\quad as\ n\rightarrow \infty ,
\end{equation*}
by applying Darboux's method \cite[Sect. 8.4]{Szego1939} to \eqref{int_repr_gener_AMP}.
%For associated Laguerre polynomials Askey and Wimp \cite[p.22]{AW1984} proved the following Mehler-Heine type formulas 
%\begin{equation}
%\lim\limits_{n\to \infty}n^{-\alpha}\mathscr{L}_n^{(\alpha)}\left(\frac{x}{n};\gamma \right)=\frac{\Gamma(\alpha)\Gamma(\gamma+1)}{\Gamma(\alpha+\gamma)}x^{-\frac{\alpha}{2}}J_{\alpha}(2\sqrt{x}), 
%\end{equation}
%when $x,\alpha >0$ and 
%\begin{equation}
%\lim\limits_{n\to \infty}\mathscr{L}_n^{(\alpha)}\left(\frac{x}{n};\gamma \right)=\Gamma(-\alpha) x^{\frac{\alpha}{2}}J_{-\alpha}(2\sqrt{x}), 
%\end{equation}
%when $\alpha <0,\ x>0$ and $\gamma\neq 0$.
\end{remark}

\section{Proofs}\label{proofs}

This section is devoted to some technical proofs. Most of them are direct and based on the series rearrangement technique and the use of some well-known transformation formulas for hypergeometric functions.
\begin{proof}[\textbf{Proof of Theorem} \ref{thm_CF_AMP}]
This can be proved by exploiting a recent result on the associated Meixner-Pollaczek polynomials $\mathscr{P}_n^{(\nu)}(x;\varphi, \gamma )$. These polynomials satisfy the recurrence relation \cite[p.2256]{Pollaczek1950}:
\begin{eqnarray}
\begin{split}\label{AMP_RR}
(n+\gamma+1)\mathscr{P}_{n+1}^{(\nu)}(x;\varphi, \gamma ) = 2[(n+\gamma+\nu)\cos\varphi +& x\sin\varphi ] \mathscr{P}_{n}^{(\nu)}(x;\varphi, \gamma ) \\
&-(n + \gamma+2\nu-1)\mathscr{P}_{n-1}^{(\nu)}(x;\varphi, \gamma )
\end{split}
\end{eqnarray}
with conditions $0<\varphi<\pi,\ 2\nu+\gamma>0$ and $\gamma\geq0$ or $0<\varphi<\pi,\ 2\nu+\gamma\geq1$ and $\nu>-1$.
By comparing the recurrence relations \eqref{AM_RR} and \eqref{AMP_RR}, one can check that the associated Meixner-Pollaczek polynomials are connected to the associated Meixner polynomials by
\begin{equation}
\mathscr{P}_n^{(\nu)}(x;\varphi, \gamma )=\frac{e^{-in\varphi}}{(\gamma +1)_n}\mathscr{M}_n(ix-\nu ;2\nu,e^{-2i\varphi } ,\gamma ).
\end{equation}
By combining the above relation and the closed-form expression of the associated Meixner-Pollaczek polynomials obtained in \cite[p.3(1.8)]{Luo2019} in terms of terminating ${}_4F_3(1)$-series, we get \eqref{AMP_4F3}. Restrictions on parameters, $c>0,\, c\neq 1$ and $\gamma +\beta >0$, can be easily derived from \eqref{coef_cond} with $n\rightarrow n+\gamma$. Now, if we put $\gamma=0$ in \eqref{AMP_4F3}, then apply the Pfaff-Kummer transformation \cite[p.33(19)]{Sr-Ma1984}:
\begin{equation}\label{Pfaff_trans}
{}_2F{}_1\left( 
\begin{array}{c}
a,b \\ 
e
\end{array}
\big|z\right)=(1-z)^{-a}{}_2F_1\left( 
\begin{array}{c}
a,e-b \\ 
e
\end{array}
\left|\frac{z}{z-1}\right.\right),
\end{equation}
for $e\neq 0,-1,-2,... , |\text{arg}(1-z)|<\pi$, with the triplet $(a,b,e)$ replaced by $(-n,-x , \beta) $ and $z=1-1/c$, we will arrive at \eqref{Meixner_poly}.
\end{proof}
%%%%%%%%%%%%%%%%%%%%%%%%%%%%%%%%%%%%%%%%%%%%%%%%%%%%%%%%%%%%%
\begin{proof}[\textbf{Proof of Theorem} \ref{thm_GF_AMP}]
Denoting the left-hand side of \eqref{gen2_AMP} by $\Upsilon(x,t)$ and substituting the expression of $\mathscr{M}_n(x;\beta , c, \gamma )$, we obtain
\begin{eqnarray*}
\begin{split}
\Upsilon(x,t)&=\sum_{n=0}^{+\infty}\frac{t^n(\gamma+1)_n}{n!}\sum_{k=0}^{n}(1-c)^k\frac{(-n)_k(\gamma+\beta+x)_k}{(\gamma +1 )_k(\gamma+\beta)_k} {}_4F_3\left( 
\begin{array}{c}
k-n,\,\gamma+\beta+x+k ,\, \gamma+\beta -1,\, \gamma\ \\ 
 \gamma+\beta +x,\, \gamma+\beta +k, \, \gamma+1+k
\end{array}
\big| 1\right)\\
&= \sum_{n=0}^{+\infty}\sum_{k=0}^{n}\sum_{j=0}^{n-k}\frac{t^n(1-c)^k}{n!j!}\frac{(-n)_k(-n-k)_j(\gamma+1)_n(\gamma+\beta+x)_k(\gamma+\beta+x+k)_j (\gamma+\beta -1)_j (\gamma)_j}{(\gamma +1 )_k(\gamma+\beta)_k(\gamma+\beta +x)_j(\gamma+\beta +k)_j(\gamma+1+k)_j}\\
&= \sum_{n=0}^{+\infty}\sum_{k=0}^{+\infty}\sum_{j=0}^{+\infty}\frac{t^{n} (t(c-1))^{k}(-t)^j}{n!j!}\frac{(\gamma+1+k+j)_{n}(\gamma+\beta+x)_k(\gamma+\beta+x+k)_j (\gamma+\beta -1)_j (\gamma)_j}{(\gamma +\beta)_k(\gamma+\beta+x)_j(\gamma+\beta+k)_j },
\end{split}
\end{eqnarray*} 
where we have used, respectively, the series transformation 
\cite[p.102(17)]{Sr-Ma1984}:
\begin{equation}\label{series_trans_3}
\sum_{n=0}^\infty\sum_{k=0}^n \sum\limits_{j=0}^{n-k}A(j,k,n)=\sum_{n=0}^\infty\sum_{k=0}^\infty \sum\limits_{j=0}^{\infty}A(j,k,n+k+j),
\end{equation}
and the identities
\begin{equation}
\frac{(-n-k-j)_k(-n-j)_j}{(n+k+j)!}=\frac{(-1)^{k+j}}{n!},
\end{equation}
\begin{equation}
\frac{(\gamma+1)_{n+k+j}}{(\gamma+1)_k(\gamma+1+k)_j}=(\gamma+1+k+j)_{n},
\end{equation}
which can be obtained from 
\begin{eqnarray}
\label{Pocha_neg_integ}
&&(-n)_k=(-1)^k\frac{n!}{(n-k)!} ,\quad k=0\leq k \leq n,\\ \label{Pocha_alter}
&&(a)_{n+m}=(a)_n(a+n)_m=(a)_m(a+m)_n, \ a\in\mathbb{C},\ n,m\in\mathbb{N}.
\end{eqnarray}
After summation over $n$ in the last expression of $\Upsilon(x,t)$, we obtain
\begin{eqnarray*}
\Upsilon(x,t)= (1-t)^{-\gamma-1}\sum_{k=0}^{+\infty}\sum_{j=0}^{+\infty}\left(\frac{t(c-1)}{1-t}\right)^{k}\frac{\left(\frac{t}{t-1}\right)^j}{j!}\frac{(\gamma+\beta+x)_k(\gamma+\beta+x+k)_j (\gamma+\beta -1)_j (\gamma)_j}{(\gamma +\beta)_k(\gamma+\beta+x)_j(\gamma+\beta+k)_j }.
\end{eqnarray*} 
We apply again the identity \eqref{Pocha_alter} to get, after simplifications, 
\begin{eqnarray*}
\Upsilon(x,t)&=& (1-t)^{-\gamma-1}\sum_{k=0}^{+\infty}\sum_{j=0}^{+\infty}\left(\frac{t(c-1)}{1-t}\right)^{k}\frac{\left(\frac{t}{t-1}\right)^j}{j!}\frac{(\gamma+\beta+x+j)_k (\gamma+\beta -1)_j (\gamma)_j}{(\gamma +\beta)_j(\gamma+\beta+j)_k}\\
&=&(1-t)^{-\gamma-1}\sum_{j=0}^{+\infty}\frac{(\gamma+\beta-1)_j(\gamma)_j}{(\gamma+\beta)_j}  {}_2F_1\left( 
\begin{array}{c}
\gamma +\beta+x+ j,\, 1\ \\ 
 \gamma+\beta+j
\end{array}
\left| \frac{t(c-1)}{1-t}\right.\right)\frac{\left(\frac{t}{t-1}\right)^j}{j!}.
\end{eqnarray*} 
Next, to ${}_2F_1$ in the above equation, we apply the Euler transformation \cite[p.33(21)]{Sr-Ma1984}:
\begin{equation}\label{Euler_trans}
{}_2F{}_1\left( 
\begin{array}{c}
a,b \\ 
e
\end{array}
\big|z\right)=(1-z)^{e-a-b}{}_2F_1\left( 
\begin{array}{c}
e-a,e-b \\ 
e
\end{array}
\big|z\right),
\end{equation}
where $e\neq 0,-1,-2,... , |\text{arg}(1-z)|<\pi$, to get
\begin{eqnarray*}
\Upsilon(x,t)=(1-t)^{x-\gamma}(1-ct)^{-x-1}\sum_{j=0}^{+\infty}\frac{(\gamma+\beta-1)_j(\gamma)_j}{(\gamma+\beta)_j}  {}_2F_1\left( 
\begin{array}{c}
-x,\, \gamma +\beta-1+ j \ \\ 
 \gamma+\beta+j
\end{array}
\left| \frac{t(c-1)}{1-t}\right.\right)\frac{\left(\frac{t}{t-1}\right)^j}{j!}.
\end{eqnarray*}
We identify the infinite series as the Appel hypergeometric function $F_1$ defined by the following double hypergeometric
series \cite[p.53(4)]{Sr-Ma1984}:
\begin{eqnarray}\label{Appel_F1}
\begin{split}
F_1\left[ \alpha ,\, \beta_1 ,\, \beta_2 ;\, \sigma ;\, x,\,y \right]&=\sum\limits_{m,n=0}^{+\infty} \frac{(\alpha)_{m+n}(\beta_1)_m(\beta_2)_n}{(\sigma)_{m+n}}\frac{x^m}{m!}\frac{y^n}{n!}\\
&=\sum\limits_{m=0}^{+\infty} \frac{(\alpha)_{m}(\beta_1)_m}{(\sigma)_{m}} {}_2F_1\left( 
\begin{array}{c}
\alpha+m , \beta_2 \\ 
\sigma+m
\end{array}
\big|y\right)\frac{x^m}{m!}, \quad \max\{|x|,|y|\}<1.
\end{split}
\end{eqnarray}
Thus, we have
\begin{eqnarray*}
\Upsilon(x,t)=(1-t)^{x-\gamma}(1-ct)^{-x-1}F_1\left[\gamma+\beta-1 ; \gamma , -x ; \gamma +\beta ; \frac{t}{t-1}, \frac{t(1-c)}{t-1} \right].
\end{eqnarray*}
To get \eqref{gen2_AMP}, it suffices to apply the transformation (see \cite[p.78]{Bailey1964}):
\begin{eqnarray}
F_1\left[ \alpha ,\, \beta_1 ,\, \beta_2 ;\, \sigma ;\, x,\,y \right]=(1-x)^{-\beta_1}(1-y)^{-\beta_2}F_1\left[ \sigma -\alpha ,\, \beta_1 ,\, \beta_2 ;\, \sigma ;\, \frac{x}{x-1},\,\frac{y}{y-1} \right]\label{F1_trans1}
\end{eqnarray}
Finally, the formula \eqref{gener2_MPP} is a particular case of \cite[Eq(9.10.13)]{askey-scheme} and it is obtained here directly by putting $\gamma=0$ in \eqref{gen2_AMP}. The proof is complete. 
\end{proof}
%%%%%%%%%%%%%%%%%%%%%%%%%%%%%%%%%%%%%%%%%%%%%%%%%%%%%%%%%%%%%%%%%%%%%%%%%%%%%%%%%%%%%%%%%%%
%\newpage
%%%%%%%%%%%%%%%%%%%%%%%%%%%%%%%%%%%%%%%%%%%%%%%%%%%%%%%%%%%%%%%%%%%%%%%%%%%%%%%%%%%%%%%%%%%%%
\begin{proof}[\textbf{Proof of Proposition} \ref{prop_GF_ACP}]
In fact, recalling the expression of \eqref{ACP_exp} and denoting the left-hand side of \eqref{gener_ACP} by $\mathscr{G}(x,t)$, we obtain
\begin{eqnarray}
\mathscr{G}(x,t)&=&\sum_{n=0}^{+\infty}\frac{t^n}{n!}\sum_{k=0}^{n}(-a)^{-k}\frac{(-n)_k(\gamma-x)_k}{(\gamma +1 )_k}
{}_3F_2\left( 
\begin{array}{c}
k-n,\,\gamma-x+k ,\, \gamma\ \\ 
 \gamma-x,\, \gamma+k+1
\end{array}
\big| 1\right)\\
&=& \sum_{n=0}^{+\infty}\sum_{k=0}^{n}\sum_{j=0}^{n-k}\frac{t^n (-a)^{-k}}{n!j!}\frac{(-n)_k(k-n)_j(\gamma-x)_k(\gamma-x+k)_j(\gamma)_j}{(\gamma +1 )_k(\gamma-x)_j(\gamma+k+1)_j }\\
&=& \sum_{n=0}^{+\infty}\sum_{k=0}^{+\infty}\sum_{j=0}^{+\infty}\frac{t^{n+k+j} (-a)^{-k}}{(n+k+j)!j!}\frac{(-n-k-j)_k(-n-j)_j(\gamma-x)_k(\gamma-x+k)_j(\gamma)_j}{(\gamma +1 )_k(\gamma-x)_j(\gamma+k+1)_j }.
\end{eqnarray} 
These calculations can be done by applying the identity \eqref{series_trans_3}. From \eqref{Pocha_neg_integ} it follows that
\begin{equation}
\frac{(-n-k-j)_k(-n-j)_j}{(n+k+j)!}=\frac{(-1)^{k+j}}{n!}.
\end{equation}
Also, from \eqref{Pocha_alter}, we have
\begin{eqnarray}
(\gamma +y)_k(\gamma +y+k)_j=(\gamma +y)_j(\gamma +y+j)_k
\end{eqnarray}
for $y=-x,\ 1$. Then, $\mathscr{G}(x,t)$ becomes
\begin{eqnarray}
\mathscr{G}(x,t)&=& e^t\sum_{k=0}^{+\infty}\sum_{j=0}^{+\infty}\left(\frac{t}{a}\right)^k\frac{  (-t)^j}{j!}\frac{(\gamma-x+j)_k(\gamma)_j}{(\gamma+1+j)_k(\gamma +1 )_j }  \nonumber \\ \label{2F1_G(x,t)}
&=& e^t\sum_{j=0}^{+\infty}\frac{  (-t)^j}{j!}\frac{(\gamma)_j}{(\gamma +1 )_j } {}_2F_1\left( 
\begin{array}{c}
\gamma-x+j ,\, 1\ \\ 
 \gamma+1+j
\end{array}
\left| \frac{t}{a}\right.\right).
\end{eqnarray}
Next, to ${}_2F_1$ in \eqref{2F1_G(x,t)} we apply the Euler transformation \eqref{Euler_trans}, to get
\begin{eqnarray}\label{G_sum_2F1}
\mathscr{G}(x,t)=e^t\left(1-\frac{t}{a}\right)^x\sum_{j=0}^{+\infty}\frac{  (-t)^j}{j!}\frac{(\gamma)_j}{(\gamma +1 )_j } {}_2F_1\left( 
\begin{array}{c}
\gamma +j,\, x+1\ \\ 
 \gamma+1+j
\end{array}
\left| \frac{t}{a}\right.\right).
\end{eqnarray}
By recognizing the Humbert confluent hypergeometric function $\Phi_1$ defined by \cite[p.58(36)]{Sr-Ma1984}:
\begin{eqnarray}\label{Phi_1F1}
\begin{split}
\Phi_1\left[\alpha_1 ; \,\lambda ;\,\alpha_2 ;\, x,\, y \right]&=\sum\limits_{m,n=0}^{+\infty} \frac{(\alpha_1)_{m+n}(\lambda)_m}{(\alpha_2)_{m+n}}\frac{x^m}{m!}\frac{y^n}{n!}\\ 
&=\sum\limits_{m=0}^{+\infty} \frac{(\alpha_1)_{m}(\lambda)_m}{(\alpha_2)_{m}} {}_1F_1\left( 
\begin{array}{c}
\alpha_1+m  \\ 
\alpha_2+m
\end{array}
\big|y\right)\frac{x^m}{m!}, \quad |x|<1,\, |y|<+\infty .
\end{split}
\end{eqnarray}
in the right-hand side of \eqref{G_sum_2F1}, we complete the proof of the Proposition \ref{prop_GF_ACP}.
\end{proof}
%%%%%%%%%%%%%%%%%%%%%%%%%%%%%%%%%%%%%%%%%%%%%%%%%%%%%%%%%%%%%%%%%%%%%%%%%%%%%%%%%%%%
\begin{proof}[\textbf{Proof of Proposition} \ref{prop_GF_ALP}]
Denote the left-hand side of \eqref{gen_ALP_1} by $\Lambda(x,t)$. Similar calculations to the ones of the proof of Proposition \ref{prop_GF_ACP} give
\begin{eqnarray}
\Lambda(x,t)&=&\sum_{n=0}^{+\infty}\frac{t^n}{n!}\sum_{k=0}^{n}\frac{(\gamma +\alpha+1 )_n(-n)_k \, x^k}{(\gamma +1 )_k(\gamma +\alpha+1 )_k}
{}_3F_2\left( 
\begin{array}{c}
k-n,\,\gamma+\alpha ,\, \gamma\ \\ 
 \gamma+\alpha+k+1,\, \gamma+k+1
\end{array}
\big| 1\right)\\
%&=& \sum_{n=0}^{+\infty}\sum_{k=0}^{n}\sum_{j=0}^{n-k}\frac{t^n x^{k}}{n!j!}\frac{(\gamma +\alpha+1 )_n(-n)_k(k-n)_j(\gamma+\alpha)_j(\gamma)_j}{(\gamma +1 )_k(\gamma +\alpha+1 )_k(\gamma+\alpha+k+1)_j(\gamma+k+1)_j }\\
%&=& \sum_{n=0}^{+\infty}\sum_{k=0}^{+\infty}\sum_{j=0}^{+\infty} \frac{t^{n+k+j} x^{k}}{(n+k+j)!j!}\frac{(\gamma +\alpha+1 )_{n+k+j}(-n-k-j)_k(-n-j)_j(\gamma+\alpha)_j(\gamma)_j}{(\gamma +1 )_k(\gamma +\alpha+1 )_k(\gamma+\alpha+k+1)_j(\gamma+k+1)_j }\\
&=& \sum_{n=0}^{+\infty}\sum_{k=0}^{+\infty}\sum_{j=0}^{+\infty} \frac{(\gamma +\alpha+1 )_{n+k+j}t^{n} }{n!}\frac{(-xt)^{k}(-t)^j(\gamma+\alpha)_j(\gamma)_j}{j!(\gamma +1 )_{k+j}(\gamma +\alpha+1 )_{k+j} }\\
&=& \sum_{n=0}^{+\infty}\sum_{k=0}^{+\infty}\sum_{j=0}^{+\infty} \frac{(\gamma +\alpha+1 +k+j)_{n}t^{n} }{n!}\frac{(-xt)^{k}(-t)^j(\gamma+\alpha)_j(\gamma)_j}{j!(\gamma +1 )_{k+j} }\\
&=& (1-t)^{-\gamma-\alpha -1} \sum_{k=0}^{+\infty}\sum_{j=0}^{+\infty} \frac{(\frac{xt}{t-1})^{k}(\frac{t}{t-1})^j(\gamma+\alpha)_j(\gamma)_j}{j!(\gamma +1 )_{k+j} }\\
&=& (1-t)^{-\gamma-\alpha -1} \sum_{j=0}^{+\infty} \frac{(\frac{t}{t-1})^j(\gamma+\alpha)_j(\gamma)_j}{j!(\gamma +1 )_{j} } {}_1F_1\left( 
\begin{array}{c}
1\ \\ 
 \gamma+1+j
\end{array}
\left| \frac{xt}{t-1}\right.\right).
\end{eqnarray} 
Next, we use the Kummer transformation \cite[p.37(7)]{Sr-Ma1984}:
\begin{equation}\label{Kummer_trans}
{}_1F_1\left( a; b; z\right)=e^z\, {}_1F_1\left( b-a; b; -z\right),
\end{equation}
to get
\begin{eqnarray}
\Lambda(x,t)=(1-t)^{-\gamma-\alpha -1}\exp\left(\frac{xt}{t-1}\right) \sum_{j=0}^{+\infty} \frac{(\frac{t}{t-1})^j(\gamma+\alpha)_j(\gamma)_j}{j!(\gamma +1 )_{j} } {}_1F_1\left( 
\begin{array}{c}
\gamma +j\ \\ 
 \gamma+1+j
\end{array}
\left| \frac{-xt}{t-1}\right.\right).
\end{eqnarray}
Identification of the summation over $j$ as the Humbert confluent hypergeometric function $\Phi_1$, then completes the proof of Proposition \ref{prop_GF_ALP}.
\end{proof}

\begin{proof}[\textbf{Proof of Proposition \ref{AMP_QF}}]
We start from the hypergeometric representation for the associated Meixner polynomials given by \eqref{m_n} which we rewrite after using the Euler transformation as follows
\begin{eqnarray}
\begin{split}\label{AMP_A,B}
\mathscr{M}_n(x;\beta ,c,\gamma)= \frac{c^{-n-x-\gamma-\beta}}{\beta-1}\left\{ (\gamma+\beta-1)_{n+1}\mathsf{A}_{n+\gamma+1}(\tilde{c})\mathsf{B}_{\gamma}(\tilde{c})-(\gamma)_{n+1} \mathsf{A}_{\gamma}(\tilde{c})\mathsf{B}_{n+\gamma+1}(\tilde{c}) \right\}
\end{split}
\end{eqnarray}
where 
\begin{eqnarray}\label{A,B}
\mathsf{A}_{\alpha}(\tilde{c}):={}_2F_1\left( 
\begin{array}{c}
x+\beta,\alpha+\beta -1 \\ 
\beta
\end{array}
\big|\tilde{c}\right) \quad \text{and} \quad \mathsf{B}_{\alpha}(\tilde{c}):={}_2F_1\left( 
\begin{array}{c}
x+1,\alpha \\ 
2-\beta
\end{array}
\big|\tilde{c}\right)
\end{eqnarray}
We apply the formula (\cite[p.48]{Magnus1996}):
\begin{eqnarray}
\begin{split}
{}_2F{}_1\left( 
\begin{array}{c}
a ,\, b \\ 
e
\end{array}
\left|z\right.\right)=\frac{\Gamma(b-a)\Gamma(e)}{\Gamma(b)\Gamma(e-a)}(1-z)^{-a} {}_2F{}_1\left( 
\begin{array}{c}
a ,\, e-b \\ 
1+a-b
\end{array}
\left|\frac{1}{1-z}\right.\right)\\
+\frac{\Gamma(e)\Gamma(a-b)}{\Gamma(a)\Gamma(e-b)}(1-z)^{-b} {}_2F{}_1\left( 
\begin{array}{c}
b ,\, e-a \\ 
1+b-a
\end{array}
\left|\frac{1}{1-z}\right.\right),
\end{split}
\end{eqnarray}
valid for $|\arg (1-z)|<\pi$, $a-b\neq \pm m, \ m=0,1,2,3,...,$ to ${}_2F_1$ in \eqref{A,B} we get
\begin{eqnarray}
&&\mathsf{A}_{\alpha}(\tilde{c})=(1-c)^{1-\beta}\left[\frac{\Gamma(\alpha-x-1)\Gamma(\beta)}{\Gamma(\alpha+\beta-1)\Gamma(-x)}c^{x+\beta}\mathsf{I}_{\alpha}(c)+\frac{\Gamma(1-\alpha+x)\Gamma(\beta)}{\Gamma(1-\alpha)\Gamma(x+\beta)}c^{\alpha+\beta-1}\mathsf{J}_{\alpha}(c)\right]\\
&&\mathsf{B}_{\alpha}(\tilde{c})=\frac{\Gamma(\alpha-x-1)\Gamma(2-\beta)}{\Gamma(\alpha)\Gamma(-x)}c^{x+1}\mathsf{I}_{\alpha}(c)+\frac{\Gamma(1-\alpha+ x)\Gamma(2-\beta)}{\Gamma(2-\beta-\alpha)\Gamma(x+1)}c^{\alpha+\beta-1}\mathsf{J}_{\alpha}(c)
\end{eqnarray}
where 
\begin{eqnarray}\label{A,B}
\mathsf{I}_{\alpha}(c):={}_2F_1\left( 
\begin{array}{c}
x+1,2-\beta-\alpha \\ 
2+x-\alpha
\end{array}
\big|c\right) \quad \text{and} \quad \mathsf{J}_{\alpha}(c):={}_2F_1\left( 
\begin{array}{c}
\alpha, 1-\beta-x\\ 
\alpha-x
\end{array}
\big|c\right)
\end{eqnarray}
By substituting $\mathsf{A}_{\alpha}(\tilde{c})$ and $\mathsf{B}_{\alpha}(\tilde{c})$ in \eqref{AMP_A,B}, we are left with sum of eight quantities containing product of $\mathsf{I}_{\alpha}(c)$ and $\mathsf{J}_{\alpha}(c)$. Four of these cancel and the remaining ones give after some algebra, in which we used essentially the relations 
\begin{equation}
\Gamma(\alpha-n)=(-1)^n\Gamma(\alpha)/(1-\alpha)_n
\end{equation}
and 
\begin{equation}
\frac{\Gamma(a)\Gamma(1-a)\Gamma(b+e)\Gamma(1-b-e)}{\Gamma(e)\Gamma(1-e)\Gamma(a+b)\Gamma(1-a-b)}+\frac{\Gamma(a)\Gamma(1-a)\Gamma(b+e)\Gamma(1-b-e)}{\Gamma(b)\Gamma(1-b)\Gamma(a-e)\Gamma(1-a+e)}=1,
\end{equation}
the final expression
\begin{eqnarray}
\begin{split}
\mathscr{M}_n(x;\beta ,c,\gamma)&=c^{-n}(\gamma -x)_{n}\ {}_2F{}_1\left( 
\begin{array}{c}
-\gamma-n ,\, \beta+x \\ 
1+x-\gamma-n
\end{array}
\left|c\right.\right) {}_2F{}_1\left( 
\begin{array}{c}
\gamma ,\, 1-\beta-x \\ 
\gamma -x
\end{array}
\left|c\right.\right)&\\
&- c\frac{(\gamma)_{n+1}(\gamma+\beta-1)_{n+1}}{(\gamma-x-1)_{n+2}}  {}_2F{}_1\left( 
\begin{array}{c}
-x ,\, \beta+\gamma+n \\ 
1+\gamma-x+n
\end{array}
\left|c\right.\right) {}_2F{}_1\left( 
\begin{array}{c}
x+1 ,\, 2-\beta-\gamma \\ 
2+x-\gamma
\end{array}
\left|c\right.\right).
&
\end{split}
\end{eqnarray}
This completes the proof.
\end{proof}
%%%%%%%%%%%%%%%%%%%%%%%%%%%%%%%%%%%%%%%%%%%%%%%%%%%%%%%%%%%%%%%%%%%%%%%%%%%%%%%%%%%%%%%%%%
\begin{proof}[\textbf{Proof of Theorem} \ref{thm_MH_ACP}]
The proof is somewhat similar to the one given in \cite{Dominici2016} for Charlier polynomials. To prove \eqref{MH-ACP}, we start from \eqref{ACP_exp2} and we write
\begin{eqnarray}
\frac{a^n}{(\gamma-x)_{n}}\mathscr{C}_n(x;a,\gamma )=a^n\sum_{k=0}^{n} \frac{(-n)_{n-k}(\gamma-x)_{n-k}}{(1)_{n-k}(\gamma-x)_{n}}
(-a)^{k-n} \ {}_3F_2\left( 
\begin{array}{c}
-k,\,\gamma ,\, k-n \ \\ 
 -n,\, \gamma-x
\end{array}
\big| 1\right).
\end{eqnarray} 
Next, we use the identities 
\begin{equation}
\frac{(\gamma-x)_{n-k}}{(\gamma-x)_{n}}=\frac{1}{(n-k+\gamma-x)_k}
\end{equation}
and
\begin{equation}
\frac{(-n)_{n-k}}{(1)_{n-k}}=(-1)^{n-k}\frac{(k+1)_{n-k}}{(1)_{n-k}}=(-1)^{n-k}\frac{(1+n-k)_{k}}{(1)_{k}}
\end{equation}
to get
\begin{eqnarray}
\frac{a^n}{(\gamma-x)_{n}}\mathscr{C}_n(x;a,\gamma )=\sum_{k=0}^{n} \frac{a^k(n-k+1)_{k}}{k!(n-k+\gamma-x)_k} \ {}_3F_2\left( 
\begin{array}{c}
-k,\,\gamma ,\, k-n \ \\ 
 -n,\, \gamma-x
\end{array}
\big| 1\right).
\end{eqnarray} 
Since, for $\gamma -x \geq 1$, we have
\begin{equation}
0<\frac{(n-k+1)_{k}}{(n-k+\gamma-x)_k}\leq 1, \quad 0\leq k\leq n
\end{equation}
and 
\begin{equation}
{}_3F_2\left( 
\begin{array}{c}
-k,\,\gamma ,\, k-n \ \\ 
 -n,\, \gamma-x
\end{array}
\big| 1\right)=\sum_{j=0}^{\infty} \frac{(-k)_j(\gamma)_j(k-n)_j}{(-n)_j(\gamma-x)_jj!}  \leq {}_2F_1\left( 
\begin{array}{c}
-k,\,\gamma  \ \\ 
  \gamma-x
\end{array}
\big| 1\right)
\end{equation}
since
\begin{equation}
0<\frac{(k-n)_j}{(-n)_j}\leq 1, \quad 0\leq j\leq min\{k, n-k\}
\end{equation}
(which can be easily deduced from the inequality $(n-k)!(n-j)!\leq n!(n-k-j)!$, for $0\leq k,j \leq n$)
then 
\begin{equation}
\frac{a^k(n-k+1)_{k}}{k!(n-k+\gamma-x)_k} \ {}_3F_2\left( 
\begin{array}{c}
-k,\,\gamma ,\, k-n \ \\ 
 -n,\, \gamma-x
\end{array}
\big| 1\right)\leq \frac{a^k}{k!}{}_2F_1\left( 
\begin{array}{c}
-k,\,\gamma  \ \\ 
  \gamma-x
\end{array}
\big| 1\right)= \frac{a^k(-x)_k}{k!(\gamma -x)_k}
\end{equation}
where we have used the identity
\begin{equation}
{}_2F_1\left( 
\begin{array}{c}
-k,\,a   \ \\ 
  c
\end{array}
\big| 1\right)=\frac{(c-a)_k}{(c)_k}.
\end{equation}
Thus, from Tanney's theorem \cite{Tannery1886}, which states that if we have $l_k \leq a_k(n) \leq u_k,\ 0\leq k\leq n$, such that $\lim\limits_{n\to \infty}a_k(n)=A_k, \ k=0,1,...,$ and that $\sum_{k=0}^{\infty}l_k, \ \sum_{j=0}^{\infty}A_k, \ \sum_{j=0}^{\infty}u_k$
are all convergent series, then, 
\begin{equation}
\lim\limits_{n\to \infty}\sum_{k=0}^{n}a_k(n)=\sum_{k=0}^{\infty}A_k,
\end{equation}
we conclude, after using the transformation ${}_1F_1(a;\ b ; x)=e^x {}_1F_1(b-a;\ b ; -x)$, that 
\begin{equation}
\lim\limits_{n\to \infty}\frac{a^n}{(\gamma-x)_{n}}\mathscr{C}_n(x;a,\gamma )=e^a {}_1F_1(\gamma;\ \gamma -x; -a),\quad x\leq \gamma -1.
\end{equation}
Dividing both sides of the above equation by $\Gamma(\gamma-x)$, we have
\begin{equation}
\lim\limits_{n\to \infty}\frac{a^n}{\Gamma(n+\gamma-x)}\mathscr{C}_n(x;a,\gamma )=\frac{e^a}{\Gamma(\gamma-x)} {}_1F_1(\gamma;\ \gamma -x; -a),\quad x\leq \gamma -1.
\end{equation}
However, since both sides of the equation are analytic in the whole complex plane (see \cite[Remark 3]{Dominici2016}),
the result follows from the principle of analytic continuation.
\end{proof}   
\subsection*{Acknowledgments} I would like to thank Professor Zouha\"ir Mouayn for his comments and invaluable suggestions that improved the presentation of this manuscript.

\end{document}